\documentclass[manuscript]{aastex}
\shorttitle{Intra-day optical variability in Mrk 501} \shortauthors{Feng et al.}

\begin{document}

\title{Search for intra-day optical variability in Mrk 501}

\author{Hai-Cheng Feng\altaffilmark{1,2,3}, H. T. Liu\altaffilmark{1,3,4}$^{\bigstar}$,
 X. L. Fan\altaffilmark{1,2,3}, Yinghe Zhao\altaffilmark{1,3,4}, J. M.
 Bai\altaffilmark{1,3,4}, Fang Wang\altaffilmark{1,2,3}, D. R. Xiong\altaffilmark{1,2,3}, and S. K. Li\altaffilmark{1,3,4}}

\altaffiltext{1} {Yunnan Observatories, Chinese Academy of Sciences, 396 Yangfangwang, Guandu District, Kunming, 650216, P. R. China}

\altaffiltext{2} {University of Chinese Academy of Sciences, Beijing 100049, P. R. China}

\altaffiltext{3} {Key Laboratory for the Structure and Evolution of Celestial Objects, Chinese Academy of Sciences, 396 Yangfangwang, Guandu District, Kunming, 650216, P. R. China}

\altaffiltext{4}{Center for Astronomical Mega-Science, Chinese Academy of Sciences, 20A Datun Road, Chaoyang District, Beijing, 100012, P. R. China}

\altaffiltext{$^{\bigstar}$}{Corresponding author: H. T. Liu, e-mail: htliu@ynao.ac.cn }

\begin{abstract}
  We present our observations of the optical intra-day variability (IDV) in $\gamma$-ray BL Lac object Mrk 501. The observations were run with
  the 1.02 m and 2.4 m optical telescopes at Yunnan Observatories from 2005 April to 2012 May. The light curve at the $R$ band on 2010 May
  15 passes both variability tests (the $F$ test and the ANOVA test). A flare within the light curve on 2010 May 15 has a magnitude change
  $\Delta m = 0.03 \pm 0.005_{\rm{stat}} \pm 0.007_{\rm{sys}}$ mag, \textbf{a darkening timescale of $\tau_{\rm{d}}=$ 26.7 minutes}, and an
  amplitude of IDV $Amp=2.9\% \pm0.7\%$. A decline \textbf{described by 11 consecutive flux measurements} within the flare can be fitted
  linearly with a Pearson's correlation coefficient $r = 0.945$ at the confidence level of $> 99.99\%$. Under the assumptions that the IDV is tightly connected to the mass of the black hole, \textbf{and that the flare duration, being two times $\tau_{\rm{d}}$,  is representative of the minimum characteristic timescale, we can derive upper bounds to the mass of the black hole}. In the case of the Kerr  black hole, the timescale of $\Delta t_{\rm{min}}^{\rm{ob}}=$ 0.89 hours gives $M_{\bullet}\la 10^{9.20} M_{\odot}$, which is consistent with measurements reported in the literature. This agreement indicates that the hypothesis about $M_{\bullet}$ and $\Delta t_{\rm{min}}^{\rm{ob}}$ is consistent with the measurements/data.
\end{abstract}

\keywords{galaxies: active --- galaxies: nuclei --- BL Lacertae objects: individual (Mrk 501)}

\section{INTRODUCTION}
   Blazars are the most violently variable subclass of active galactic nuclei (AGNs), which include BL Lacertae objects (BL Lacs) and flat spectrum
   radio quasars (FSRQs). They are characterized by rapid and strong variability at all wavelengths of the whole electromagnetic spectrum, strong
   and variable polarization from radio to optical bands, non-thermal emission dominating at all wavelengths. Blazars usually exhibit the core-dominated radio shape. These extreme properties are generally believed to be generated from a relativistic jet with a viewing angle less
   than $10^{\circ}$ \citep[e.g.,][]{BK79,UP95}. The broadband spectral energy distributions (SEDs) of blazars usually show a double-peak profile.
   The low-energy component spans the IR-optical-UV bands and the high-energy component extends to GeV/TeV gamma-ray bands \citep[e.g.,][]{Gh98,Ab10a}. The first peak is generally interpreted as the synchrotron radiation of relativistic electrons in the jet. The second
   peak is generally believed to be generated by the inverse-Compton scattering of the same electron distribution responsible for the synchrotron radiation \citep[e.g.,][]{DS93,Bo07,Ne12}.

   Photometric technique is a powerful tool to investigate the nature of blazars. The variability including amplitude, duty cycles, timescales and
   so on, could help us to study the structure, dynamics, radiation mechanism and even phase of blazars \citep[e.g.,][]{Ci03,Ci07,Gu08a,Da15,Xi15}.
   Previous observations show that blazars exhibit variability on diverse timescales, which can be broadly divided into three classes: intra-day
   variability (IDV) or micro-variability, short-term variability (STV), and long-term variability (LTV). For IDVs, the timescales range from a few
   minutes to several hours, and the variability of flux changes by a few tenths of magnitude \citep[e.g.,][]{WW95}. STVs and LTVs have timescales
   of days to months and months to years, respectively \citep[e.g.,][]{Fa05,Fa09,Gu08b,Da15,Xi15}. Over the last three decades, a great mount of
   work for variability of blazars has been reported at different timescales in different bands \citep[e.g.,][]{Mi89,Ba98,Ca00,Xi90,Xi91a,Xi91b,Xi92,Xi99,Xi01,Xi02b,Xi02c,Xi05a,Fa05,Fa09,Ah07,Ci03,Ci07,Gu04,Gu08a,Gu08b,Da15,Xi15,Ah16}.
   The IDVs of the beamed emission of jets can be used to constrain the central black hole masses of blazars \citep{Li15}.

   Mrk 501 is a typical nearby BL Lac object (at redshift $z=0.034$), one of the nearest Jy BL Lac objects  \citep{St93,Ko03} and one of the
   brightest extragalactic sources in the X-ray/TeV sky \citep{Ab11}. It has attracted much attention. Since the first detection by Whipple
   observatory \citep{Qu96}, Mrk 501 was the second extragalactic object identified as a very high energy (VHE) $\gamma$-ray emitter \citep{Ko03,Ab11,Xi15}. In the second year since it was discovered, Mrk 501 went into a surprisingly high activity and strong variability
   state. Fluxes over 10 times brighter than 10 Crab were reported in several groups \citep{Ca97,Sa98,Ah99a,Ah99b}. After the outburst state,
   the mean VHE $\gamma$-rays dropped by an order of magnitude during 1998--1999 \citep{Ah01}. In 2005, an interesting phenomenon,
   the VHE $\gamma$-ray flux varies on a timescale of minutes, was detected \citep{Al07}. The recent report about a multi-wavelength
   study in 2009 shows fast variability with $\sim$ 15 minute doubling times in the VHE gamma rays \citep{Ah16}. The variability of Mrk 501
   on the whole electromagnetic wavelengths has been extensively studied \citep[e.g.,][]{Xi99,Xi01,Gu08b,Ro09,Gu12,Al15,Sh15,Wi15,Xi15,Ah16}.
   The IDVs of Mrk 501 have been reported over the entire electromagnetic spectrum \citep[e.g.,][]{Al07,Gu08b,Sh15,Xi15}. Most of previous works
   were on timescales of tens to one hundred minutes for the optical IDVs. A rapid X-ray flare from Mrk 501 detected with the Rossi X-Ray Timing Explorer in 1998 shows the 60\% increase in 200 seconds ($\sim$ 3 minutes) \citep{Ca00}. \citet{Xi99} reported a $R$ band flare on a timescale
   of 105 minutes. \citet{Xi15} found a $R$ band flare around 106 minutes. Compared to the X-ray and VHE flares on timescales of minutes,
   the optical IDVs are on timescales of hours. Shorter optical IDVs may be expected.

   In this paper, we present observations and investigate the variability in $B$, $I$, $R$, and $V$ bands from 2005 to 2012. To find out the fastest
   optical IDV, most exposure times are less than 3 minutes. The structure of this paper is as follows: in section 2 we present the observations and
   data reduction; the results of variability are exhibited in section 3; section 4 is for discussion.

\section{OBSERVATIONS AND DATA REDUCTION}

   The optical monitoring program of Mrk 501 was carried out with two telescopes: the 1.02 m and 2.4 m optical telescopes at Yunnan Observatories, China. The 1.02 m telescope is located at Kunming, China. From 2006 to 2009, the telescope equipped with a RAC CCD (1024 pixels $\times$ 1024 pixels) at $f$/13.3 Cassegrain focus, and the entire CCD chip covered $\sim$6.5 $\times$ 6.5 arcmin$^{2}$. So, the projected angle on the sky of
   each pixel corresponded to 0.38 arcsec in both dimensions. The readout noise and gain were 3.9 electrons and 4.0 electrons/ADU, respectively.
   After 2009, the telescope was equipped with an Andor AW436 CCD (2048 pixels $\times$ 2048 pixels) camera at the $f$/13.3 Cassergrain focus.
   The field of view of the CCD is $\sim$7.3 $\times$ 7.3 arcmin$^{2}$, and the pixel scale is 0.21 arcsec in both dimensions. The readout noise and
   gain were 6.33 electrons and 2.0 electrons/ADU, respectively \citep[e.g.,][]{Xi02b,Gu08a,Da15}. The 2.4 m telescope is located at Lijiang, China. The telescope was equipped with an Princeton Instruments VersArray1300B CCD (1300 pixels $\times$ 1340 pixels) camera at f/8 Cassegrain focus,
   and the entire CCD chip covered $\sim$4.40 $\times$ 4.48 arcmin$^{2}$. The readout noise and gain were 6.5 electrons and 1.1 electrons/ADU, respectively. For both of the two telescopes, we selected standard Johnson broadband filters to carry out the observations in the $B$, $V$, $R$,
   and $I$ bands.

   During our monitoring program from 2005 to 2012, 1532 CCD frames were obtained in 50 nights. For most nights, the exposure times are
   40--400 seconds for 1.02 m telescope and 10--50 seconds for 2.4 m telescope. The complete observation log is listed in Table 1. For each
   image, the standard stars and object were always in the same field. The standard stars considered do not change, so the brightness of the
   object was obtained from the standard stars \citep[e.g.,][]{Ba98,Fa14,Zh04,Zh08}. Because star 1 is the brightest of all comparison stars and
   the magnitudes have been measured in all bands in other works \citep{Vi98,FT96}, it was selected to calculate the object magnitude. However,
   there are some uncertainties which may cause the standard stars to change, so we chose another standard star to characterize the change,
   and we used the standard deviation of the two comparison stars [$\sigma$(star1-star6)] as the error of photometry. Star 6, closest to the object
   and with all four bands measured at the same time with star 1, is used as the second standard star. The standard deviation of the differential instrumental magnitude of star1-star6 is given in Table 1. The IDV of the target object was investigated using two statistical methods: the $F$
   test \citep[e.g.,][]{de10,Jo11,Go12,Hu14,AG15} and the one-way analysis of variance \citep[ANOVA;][]{de98,de10,Hu14}. The value of $F$ test in
   our work is calculated as:
   \begin{equation}
     F = \frac{Var(BL-star1)}{Var(star1-star6)},
   \end{equation}
   where BL, star1, and star6 are the magnitudes of BL Lac object, star1, and star6, respectively. Var(BL-star1) and Var(star1-star6) are the variances
   of differential instrumental magnitudes for 'BL-star1' and 'star1-star6', respectively. The critical value of the $F$ test can be compared with the
   $F$ value for which the significance level was set as 0.01 in $F$-statistics. ANOVA is a powerful tool to detect IDVs, and does not rely on error measurements. We divided data points into groups. Each group contains three data points, and if the last group has less than three data points,
   we would merger them with the previous group. The ANOVA critical value can be obtained from $F^{\alpha}_{\nu1,\nu2}$ in the F-statistics,
   where $\nu1$ ($\nu1 = k-1$, and $k$ is the number of groups) is the degree of freedom between groups and $\nu2$ ($\nu2 = N - k$, and $N$
   is the number of measurements) is the degree of freedom within these groups, and $\alpha$ is the significance level \citep[e.g.,][]{Hu14}. When calculating the values of $F$ and ANOVA, we only used the light curves with observational data points $\geq$ 9 in a night, and only the light
   curves satisfied with both criteria were considered to be variable.  Table 1 shows the results of $F$ test and ANOVA for all the observable nights satisfying the criteria.

   All of the photometric data were reduced using the standard procedure in the Image Reduction and Analysis Facility (IRAF) software. For
   each night, we combined all the bias frames and then obtained a master bias. All of the object image frames and flat-field image frames
   were subtracted by the master bias. Then we generated the master flat-field for each band by taking the median of all flat-field image frames
   for each band. After the flat-field correction, aperture photometry was performed by the APPHOT task. Because Mrk 501 is an extended source
   and the standard stars are point like sources, we used two different criteria to determine the aperture size for the object and comparison stars.
   For the standard stars, we found that the best signal-to-noise ratio was obtained with the aperture radius of 1.2 full width at half maximum
   (FWHM), i.e. a dynamic aperture, which is generally applied to the point like sources consisting of AGNs and their host galaxies, and is large
   enough to cover the hosts. For extended sources, such as nearby AGNs, the change of seeing may affect the dynamic aperture and then the contribution of the hosts. If the host is considered to be constant for Mrk 501, the photometric results could reflect a combined contribution
   of the aperture effect and the intrinsic variations of AGN. We chose 16 different aperture radii among 1.2--8.0 FWHMs, and the variability of
   Mrk 501 is similar to each other when the apertures are greater than or equal to than 3 FWHMs (see the upper left panel in Figure 1). The light
   curves with apertures of 6.0--8.0 FWHMs have almost the same outline, i.e., the light curve profile is stable as the aperture is enough large.
   It seems that there is an arc dip on 2010 May 17 lasting for about 90 minutes. However, the aperture is dynamic with the FWHM of standard
   star, and the light curve profile may be biased by the FWHM variability. The lower left panel in Figure 1 shows the FWHM variability. This FWHM variability is very similar to those of the light curves in 1.2--2.5 FWHMs, especially the arc dip. The smaller FWHM results in the smaller aperture,
   and the fewer host contribution. This will result in the darkening of Mrk 501. A flare in the $I$ band on 2009 May 11 has a large magnitude change
   of $\Delta I$ = $0^{m}.$28 and a symmetrical profile, when the dynamical aperture is used. However, the corresponding FWHMs nearly have the same variability as the light curve of Mrk 501 (see Figure 1). Thus, the dynamical aperture significantly influences the variability results of the photometry due to the FWHM variability, and the fixed aperture is used to perform photometry. The fixed aperture can avoid this host influence
   on the measured magnitude variability of Mrk 501. In order to contain the most contribution of AGN and avoid the effect of background, we used
   3.5 arcsec aperture radius, which is more than 1.5 $\times$ typical FWHM for most of images.

   Though the fixed aperture can avoid the host influence of the dynamic aperture on the variability of Mrk 501, the seeing effect influences
   the FWHM, and then the fraction of the host light contained by the fixed aperture. Because the profile of the host galaxy is not flat \citet{Ni07},
   the different point spread functions influence the contribution of the host in a fixed aperture. However, this influence is much smaller than that
   of the dynamic aperture mentioned above. For this host, the larger seeing results in the larger point spread function that leads to the smaller faction of the host light contained by the fixed aperture. The increasing FWHM will generate the darkening of measured magnitude. The observational data and the fit results confirm the deduced FWHM--magnitude relation (see Figure 2). From 2010 May 15 to 2010 May 18, four night observations are used to test this relation. Also, there is a similar relation for these four night data. Thus, the FWHM--magnitude relation is corrected for each night data. The host contribution is subtracted according to the method of \citet{Ni07} for the data on 2010 May 18, and the FWHM--magnitude relation appears more significant. The host-corrected magnitudes depend on point spread function or seeing (FWHM). This FWHM--magnitude relation is corrected to get the final result for each night data. Figure 3 shows the light curves in the $I$ band on 2009 May 11 for a fixed aperture without and with correcting the FWHM--magnitude relation. Comparison of the light curves in Figure 3 shows that the seeing effect is improved. The IDV seems to exist in the corrected light curve in Figure 3. The $F$ test or the ANOVA test show six possible IDVs, and only the IDV on 2010 May 15 is confirmed by the two tests (see Table 1). \textbf{Two} interesting light curves with the possible IDVs are presented in Figure 4.

\section{RESULTS}
    Table 1 shows six nights for which the the light curve gave positive in the $F$ or ANOVA tests, from which only one night, 2010 May 15, gave
    positive in both tests. As an example, Figure 4 shows two nights, 2010 May 15 and 17.  These light curves have a peak-to-peak flux change
    of $\Delta R \sim 0.03$ mag, while the observational error estimated by the standard deviation from the magnitude differences between standard star 1 and star 6 is $\sigma_{\rm{R}} =0.005$. Therefore, the peak-to-peak flux change $\Delta R \sim 6\sigma_{\rm{R}}$. We followed the method
    in \citet{Du14} to determine the \textbf{systematic} error of the fluxes. First, we used a median filter to smooth the light curve, and then subtracted
    it from the original light curve. Second, we calculated the standard deviation of the residual light curve, and adopted this deviation as an estimate
    of the \textbf{systematic} uncertainty. \textbf{The value estimated with this method is 0.007 mag. If this systematic error is combined quadratically with the statistical error, we find a total uncertainty in the flux variation of $\sigma_T$=0.009, which leads to a peak-to-peak flux change of
    $\Delta R \sim 3.3 \sigma_T$.} The detection success rate of the optical IDVs seems to be very low for Mrk 501. This may be due to the intrinsic
    weak IDVs of Mrk 501 and/or the relatively brighter host galaxy. For the large amplitude variations of BL Lacertae objects, an effective variation
    on a short timescale (from a few hundred seconds to several hours) requires that the amplitude of optical variability must be more than 5$\sigma$, where $\sigma$ is the maximum total observational rms error \citep{Xi90,Xi92,Xi04}. The rms error used in \citet{Xi04} is the same as the standard deviation of the two comparison stars we used in this paper. The ratio of $\Delta m / \sigma>5$ \textbf{(where $\sigma$ relates to the statistical uncertainty)} was used as a necessary criterion rather than a sufficient and necessary condition for optical variability. Though, the light curves on 2010 May 15 and 17 match the criterion $\Delta m > 5\sigma$, they do not necessarily have IDVs \textbf{because of the systematic uncertainties mentioned above.}

   The long-term light curves are presented in Figure 5 for our observational data.  The larger amplitudes of variability are found in the long-term
   light curves. The continuous observations in the light curves appear to be a cluster, which may show a larger amplitude of variability. A variability amplitude of $\Delta B \sim 0.8$ mag is around MJD=54900 for the $B$ band. We find a variability amplitude of $\Delta I \sim 0.7$ mag for
   the $I$ band around MJD=55250. The long-term light curve in the $R$ band is similar to that in the $I$ band. However, the $R$ band FWHMs are larger than 5.3 arcseconds on 2006 April 3, which may be due a worse observational condition. The $I$ band FWHMs are smaller than those
   of the $R$ band on 2006 April 3, and the $I$ band observations were completed before the $R$ band observations.  This may result in a worse photometric magnitude in the $R$ band, darker than 18 mag (see Figure 5). The other data points of the $R$ band are not particularly surprising. There is a variability amplitude of $\Delta R \sim 0.7$ mag around MJD=55250.  For the $V$ band around MJD=54200, a variability is found with an amplitude of $\Delta V\sim 0.7$ mag. Also, a variability is found with an amplitude of $\Delta V\sim 0.7$ mag around MJD=56000.  These variations with large amplitudes of $\Delta m \sim$0.7--0.8 mag have durations within $\sim$ 100 days. Also, the similar timescale variations are found in Seyfert galaxies, e.g., NGC 5548 \citep{Ul97}. These timescales within $\sim$ 100 days may be from the lighthouse effect of
   a jet, where the jet precession will result in the forward beaming of the emission \citep[e.g.,][]{CK92,GW92}.

\section{DISCUSSION}
   Except for the jet origin of the optical variations, there is an alternative way to explain the optical variability and the IDVs or micro-variability
   in BL Lac objects, i.e. the optical IDVs are likely from accretion disks \citep[e.g.,][]{Gu12,Ag16}. The accretion disk instabilities are able to explain
   some of the phenomena seen in the optical--X-ray bands, but cannot explain the radio IDVs \citep[e.g.,][]{WW95}. The latest research on BL Lac object PKS 0735+178 shows that the blazar variability on timescales from years down to hours--i.e., both the long-term large amplitude variability and the micro-variability--is generated by the underlying single stochastic process (at the radio and optical bands), or a linear superposition of
   such processes (in the gamma-ray regime), within a highly non-uniform portion of the flow extending from the jet base up to the $\la$ pc-scale distances \citep{Go17}. \citet{Im11} searched for short timescale variability, and identified an interesting event in the $J$ band with a duration
   of $\sim$ 25 minutes for BL Lac object PKS 0537-441. In both the low and high states, the emission appears to be dominated by the jet, and no evidence of a thermal component is apparent for PKS 0537-441. The spectral energy distributions of PKS 0537-441 are interpreted in terms of the synchrotron and inverse Compton mechanisms within a jet, where the plasma radiates via internal shocks and the dissipation depends on the distance of the emitting region from the central engine \citep{Pi07}. The optical emission of Mrk 501 is neither the thermal component from accretion disk nor the nonthermal component from the jet \citep{Ah16}. The optical emission is dominated by the host galaxy, and the UV
   emission is from the jet for Mrk 501 \citep[e.g.,][]{Ah16}. Thus, it is not possible that the optical IDVs are from accretion disk for Mrk 501.

   The IDVs discussed here may be directly related to shock processes in a jet. The shock-in-jet model, the most frequently cited model used
   to explain the IDVs, is based on a relativistic shock propagating down a jet and interacting with a highly non-uniform portion in the jet flow \citep[e.g.,][and references therein]{Na12,Su12,Ma14,Sa15}. The featureless optical continuum is the typical characteristic of BL Lac objects,
   and quasars show many strong broad emission lines. The broad-line region (BLR) seems to not be in BL Lac objects \citep[e.g.,][]{UP95}.
   Broad emission lines were observed only in a few BL Lac objects \citep[e.g.,][]{Ca99,Ce97}. Thus, the broad emission lines seem to be weak
   so that the broad lines were rarely observed in BL Lac objects. The accretion rates are very low for BL Lac objects \citep[e.g.,][]{Ca02,Xu09}.
   The absence of broad emission lines in most of BL Lac objects may be due to the very weak emission of accretion disk. The BLR clouds are
   right there though observations have not detected the broad lines in most of BL Lac objects. Thus, the optical variability in BL Lac objects
   is not dominated by the emission of accretion disk. On the contrary, the relativistic jets are likely dominating the optical variability. Though
   a minority of BL Lac objects have the broad emission lines, they have Full width at half maximum $v_{\rm{FWHM}}\sim$ 1300--5500 $\rm{km \/\ s^{-1}}$ \citep{Ca04}. Their widths are very similar to those of Seyfert galaxies. This similarity indicates a fundamental difference in accretion
   rate between these two kinds of AGNs. This accretion rate difference implies that BL Lac objects do not have the same origin as Seyfert galaxies
   have an accretion disk origin of the optical variability. Mrk 501 has a featureless continuum, and then its optical IDVs do not have the accretion
   disk origin.

   The variability timescales were defined in different ways. The doubling timescale is usually used to estimate the variability timescale of the
   large amplitude change in gamma rays \citep[e.g.,][]{Fa99}. A fast variability in the VHE gamma rays shows doubling times $\sim$ 15 minutes
   for Mrk 501 \citep{Ah16}. But the fastest variability observed on Mrk 501 at VHE is $\sim$ 2 minutes \citep{Al07}. These VHE gamma rays are
   from the relativistic jets. \textbf{These variability timescales of $\sim$ 2--15 minutes could give upper limits to the diameter sizes of gamma-ray emitting regions}, $D_{\gamma} \la c \Delta t_{\rm{min}}^{\rm{ob}} \delta /(1+z)$, where $c$ is the speed of light, $\Delta t_{\rm{min}}^{\rm{ob}}$
   is the minimum variability timescale observed, and $\delta$ is the \textbf{corresponding} Doppler factor. In general, the lower energy bands will
   show the smaller variability amplitudes for Mrk 501 \citep{Ah16}. The early common definition of the variability timescale is $\tau =F /|\Delta F / \Delta t|$, and the more conservative approach is $\tau =|\Delta t / \Delta \ln F|$, where $F$ is the flux and $\Delta F$ is the variability amplitude
   on the timescale $\Delta t$ \citep[e.g.,][]{WW95}. These definitions were used to the variability with a large amplitude change, e.g., $\ga$ 0.1 mag. Another choice is the interval between the adjacent local minima at the adjacent valleys in the light curve, respectively (i.e., the flare duration),
   or two times the interval between the adjacent local minimum and maximum at the adjacent valley and peak, respectively, if the flare is not complete. As a shock passes through the emitting region (knot or blob) in the jet, this passage will generate a flare with the duration comparable
   to the passing timescale of the shock. In the case, the variability timescale corresponds to the flare duration, i.e., the interval between the adjacent local minima at the adjacent valleys in the light curve. Another possibility is that as the blob-like emitting region becomes optical thin, it generates
   a flare with a duration about equal to the light crossing time of the emitting region. These definitions should locate by hand the corresponding
   data points in the short-term light curve.

   The minimum timescale was determined by hand for the short-term light curve \citep[e.g.,][]{Xi90,Xi91a,Xi91b,Xi92,Xi99,Xi01,Xi02c,Xi05a}.
   In an analogous fashion, we searched possible flares in the light curves on 2010 May 15 and 17. There is a darkening in the light curve on
   2010 May 15 (see Figure 6). This darkening consists of 11 data points, has $\Delta R=0.030\pm 0.005$ mag and lasts for 26.7 minutes. The corresponding rate of the change is 0.067 mag/hr. The 11 data points can be fitted linearly with a Pearson's correlation coefficient $r = 0.945$
   at the confidence level of more than 99.99 per cent. Thus, the random fluctuation origin of this darkening can be eliminated at a high
   confidence level. \textbf{There seems no significant adjacent rising phase before this darkening (see Figure 6). We do not get the total
   duration of the flare from the sum of the rising and darkening timescales. The two times of the darkening timescale may be assumed as
   a representative value for the duration, 0.89 hours.} According to the necessary condition of optical variability $\Delta m > 5\sigma$ used
   in \citet{Xi90,Xi92,Xi04}, an IDV might be indicated for this flare, because the uncertainty $\sigma$ does not take into account the
   systematic uncertainty. The variability amplitude can be calculated by $Amp = \sqrt{(A_{\rm{max}} - A_{\rm{min}})^{2} - 2\sigma^{2}}$
   \citep{HW96}, where $A_{\rm{max}}$ and $A_{\rm{min}}$ are the maximum and minimum magnitudes of the light curve being considered, respectively, and $\sigma$ is the measurement error. There are $Amp=4.3\% \pm 0.7\%$ for the light curve on 2010 May 15, and $Amp=2.9\%
   \pm 0.7\%$ for the flare mentioned above. The auto-correlation function method could search for the characteristic timescale of the large
   amplitude variability in the long-term light curve \citep[e.g.,][]{Ne96,Gi99,Li08}. The $F$ test and the ANOVA test indicate the IDV in the light
   curve on 2010 May 15, but cannot give the IDV details, e.g, timescale. The structure function (SF), introduced by \citet{Si85}, \textbf{has been employed} to quantify the characteristic timescale for a light curve with confirmed variability (e.g., meeting some tests) \citep[e.g.,][]{Ab10b,Da15}. The first-order SF is defined as
   \begin{equation}
     SF^{(1)}(\Delta t)=\frac{1}{N}\sum^{N}_{i=1}[m(t_{i})-m(t_{i}+\Delta t)]^2,
   \end{equation}
   where $m(t)$ is the magnitude at time $t$, and $\Delta t$ is the time separation. The characteristic timescale in a light curve is indicated
   by a local maximum of the SF. The first local maximum was used, i.e., the one with the shortest time, in the case of multiple local maxima
   in the SF \citep[e.g.,][]{Da15}. Figure 7 displays the first-order SF for the light curve on 2010 May 15. However, the SF cannot give "a real break
   or peak" (a large break or peak), i.e., a robust variability characteristic timescale. \textbf{The SF is not able to determine the variability timescale}
   of the micro-amplitude flare shown in Figure 6. The SF increases with the timescale, which implies that there is more power of variability at the longer timescales (which is a typical characteristic in blazars and AGNs in general).

   The IDV is an intrinsic phenomenon and tightly constrains the sizes of the emitting regions in blazars \citep{WW95}. The timescales of variations
   in the optical band might have an underlying connection with the black hole masses of the central engines in blazars. The minimum timescales
   of variability were used to determine the masses of the central black holes in AGNs \citep[e.g.,][]{Ab82,Mi89,Xi02a,Xi05b,Li15}. \citet{Ab82} and \citet{Xi02a} determined the black hole masses $M_{\bullet}$ for non-blazar-like AGNs or some AGNs with a weaker blazar emission component
   in fluxes relative to an accretion disk emission component, based on the assumption that an accretion disk is surrounding a supermassive black
   hole, and the optical flux variations are from the accretion disk. \citet{Li15} proposed a sophisticated model to constrain $M_{\bullet}$ using the
   rapid variations for blazars. The model is suitable to constrain $M_{\bullet}$ in blazars using the minimum timescale $\Delta t_{\rm{min}}^{\rm{ob}}$ of variations of the beamed emission from the relativistic jets. \citet{Li15} gave
   \begin{mathletters}
   \begin{eqnarray}
     M_{\bullet}\la 5.09\times 10^4 \frac{\delta \Delta t_{\rm{min}}^{\rm{ob}}}{1+z}M_{\odot} \/\ \/\ (\/\ j\sim 1),\\
    M_{\bullet}\la 1.70 \times 10^4 \frac{\delta \Delta t_{\rm{min}}^{\rm{ob}}}{1+z}M_{\odot} \/\ \/\ (\/\ j=0),
   \end{eqnarray}
   \end{mathletters}
   where $\Delta t_{\rm{min}}^{\rm{ob}}$ is in units of seconds, and $j=J/J_{\rm{max}}$ is the dimensionless spin parameter of a black hole with
   the maximum possible angular momentum $J_{\rm{max}}=G M_{\rm{\bullet}}^2/c$  and $G$ being the gravitational constant. For the light curve
   of Mrk 501 on 2010 May 15, the optical IDV has a darkening timescale of $\tau_{\rm{d}}=$ 26.7 minutes with a micro-amplitude of $\Delta m=0.03$ mag. \textbf{We made the assumption of considering the duration of 0.89 hours, two times $\tau_{\rm{d}}$, as a representative value for the variability timescale to be used in formulae (3a) and (3b). Other studies used other prescriptions to estimate a variability timescale (like flux-doubling times, or characteristic times in SF), which cannot be used with the optical data reported in this paper.} The optical--$\gamma$-ray emission is mostly the Doppler boosted emission of jets for $\gamma$-ray blazars \citep{Gh98}. A value of $\delta \sim 10$ was adopted for GeV gamma-ray blazars \citep{Gh10}. As in \citet{Li15}, $\delta = 10$ is taken to estimate $M_{\bullet}$ with formulas (3) for Mrk 501. In the case of $\Delta t_{\rm{min}}^{\rm{ob}}=$ 0.89 hours, we have $M_{\bullet} \la 10^{9.20}\/\ M_{\odot}$ for the Kerr black hole [formula (3a)] and $M_{\bullet}\la 10^{8.72}\/\ M_{\odot}$ for the Schwarzchild black hole [formula (3b)]. The flare duration of 0.89 hours gives $M_{\bullet}\la 10^{8.72}$--$10^{9.20} M_{\odot}$. \citet{Ba03} and \citet{Fa02} measured the stellar velocity dispersion and estimated $M_{\bullet}= 10^{9.21\pm 0.13} M_{\odot}$ and $M_{\bullet}=10^{8.93\pm 0.21} M_{\odot}$, respectively, by the black hole mass--stellar velocity dispersion relation. \citet{Gh10} reported a mass $M_{\bullet}=10^{8.84} M_{\odot}$. The mass upper limit of $M_{\bullet}\la10^{9.20}\/\ M_{\odot}$ is basically consistent with the mass estimates $M_{\bullet}= 10^{8.93 \pm 0.21} M_{\odot}$, $10^{9.21\pm 0.13} M_{\odot}$, and $10^{8.84} M_{\odot}$. \citet{Ri03} showed the large uncertainties in the determination of the black hole mass of Mrk 501, ranging from $\sim 10^{7.8} M_{\odot}$ to $10^{8.7} M_{\odot}$, which are \textbf{also} consistent with the mass upper limits. We made the assumption that IDV is tightly connected to the black hole, in order to set constraints to the
   black hole mass using the method proposed in \citet{Li15}. The derived upper bounds indicate that these two hypotheses are consistent with the measurements/data. The presence of systematic uncertainties caused by poor weather conditions, telescope tracking inaccuracies, etc, will mask
   the optical IDV finding for Mrk 501. Based on the method used in \citet{Du14}, we estimated the systematic uncertainty from the standard deviation of the residuals, which were obtained by subtracting the median-smoothed light curve from the original light curve. This gives a systematic uncertainty $\thickapprox 0.007$ on 2010 May 15.  So, the IDV on 2010 May 15 is not a very robust claim because of the presence of a systematic uncertainty at the level of $\sim 1\%$ \textbf{that may affect the flux variations that were measured, which are at the level of $\sim 3\%$}. Higher accuracy observations with large aperture telescopes are needed in future.

\acknowledgements {We are grateful to the anonymous referee for constructive comments leading to significant improvement of this paper.
We thank the financial supports of the National Natural Science Foundation of China (NSFC; grants No. 11273052 and U1431228), and the Youth Innovation Promotion Association, CAS. We also acknowledge the support of the staff of the Lijiang 2.4m telescope. Funding for this telescope
has been provided by CAS and the People's Government of Yunnan Province. }

\clearpage

\begin{deluxetable}{lcccccccc}
  \tablecolumns{9}
  \setlength{\tabcolsep}{3pt}
  \tablewidth{0pc}
  \tablecaption{Observation log and results of IDV observations of Mrk 501}
  \tabletypesize{\scriptsize}
  \tablehead{
  \colhead{Date}                                             &
  \colhead{Filters}                                          &
  \colhead{N}                                                &
  \colhead{$\sigma$(star1-star6)}                            &
  \colhead{$F$}                                             &
  \colhead{$F$(99)}                                          &
  \colhead{ANOVA}                                     &
  \colhead{ANOVA(99)}                                        &
  \colhead{Telescope}  \\

\colhead{(1)}&\colhead{(2)}&\colhead{(3)}&\colhead{(4)}&\colhead{(5)}&\colhead{(6)}&
\colhead{(7)}&\colhead{(8)}&\colhead{(9)}
}
\startdata

2005 Apr 05	&	B	&	4	&	0.074	&		&		&		&		&	1.02	\\
	&	I	&	5	&	0.015	&		&		&		&		&	1.02	\\
	&	R	&	4	&	0.01	&		&		&		&		&	1.02	\\
	&	V	&	5	&	0.015	&		&		&		&		&	1.02	\\
2005 Apr 06	&	B	&	1	&		&		&		&		&		&	1.02	\\
	&	I	&	9	&	0.013	&	0.23	&	6.03	&	3.37	&	10.92	&	1.02	\\
	&	R	&	10	&	0.018	&	0.19	&	5.35	&	0.41	&	9.55	&	1.02	\\
	&	V	&	9	&	0.033	&	0.02	&	6.03	&	0.29	&	10.92	&	1.02	\\
2006 Apr 03	&	I	&	4	&	0.054	&		&		&		&		&	1.02	\\
	&	R	&	4	&	0.024	&		&		&		&		&	1.02	\\
2007 Mar 27	&	I	&	39	&	0.012	&	0.48	&	2.16	&	3.45	&	2.96	&	1.02	\\
	&	V	&	28	&	0.076	&	0.03	&	2.51	&	1.13	&	3.63	&	1.02	\\
2007 Mar 28	&	I	&	5	&	0.008	&		&		&		&		&	1.02	\\
	&	V	&	3	&	0.01	&		&		&		&		&	1.02	\\
2007 Mar 29	&	I	&	8	&	0.005	&		&		&		&		&	1.02	\\
	&	V	&	4	&	0.023	&		&		&		&		&	1.02	\\
2007 Mar 30	&	I	&	8	&	0.006	&		&		&		&		&	1.02	\\
	&	V	&	5	&	0.027	&		&		&		&		&	1.02	\\
2007 Apr 15	&	I	&	13	&	0.015	&	0.46	&	4.16	&	2.81	&	6.99	&	1.02	\\
2007 Apr 22	&	B	&	2	&	0.013	&		&		&		&		&	1.02	\\
	&	I	&	10	&	0.006	&	0.63	&	5.35	&	0.31	&	9.55	&	1.02	\\
	&	V	&	10	&	0.04	&	0.22	&	5.35	&	1.54	&	9.55	&	1.02	\\
2007 Apr 23	&	I	&	4	&	0.005	&		&		&		&		&	1.02	\\
2007 Apr 24	&	I	&	15	&	0.007	&	2.42	&	3.7	&	0.54	&	5.99	&	1.02	\\
	&	V	&	14	&	0.052	&	0.21	&	3.91	&	6.22	&	6.55	&	1.02	\\
2007 May 08	&	I	&	10	&	0.012	&	1.21	&	5.35	&	0.42	&	9.55	&	1.02	\\
2007 May 09	&	B	&	1	&		&		&		&		&		&	1.02	\\
	&	I	&	9	&	0.009	&	0.58	&	6.03	&	1.68	&	10.92	&	1.02	\\
	&	V	&	5	&	0.032	&		&		&		&		&	1.02	\\
2008 May 06	&	I	&	4	&	0.037	&		&		&		&		&	1.02	\\
2008 May 07	&	I	&	87	&	0.009	&	1.05	&	1.69	&	1.81	&	2.08	&	1.02	\\
2008 May 08	&	I	&	82	&	0.008	&	1.38	&	1.69	&	3.08	&	2.11	&	1.02	\\
2009 Mar 21	&	B	&	2	&	0.001	&		&		&		&		&	2.4	\\
	&	R	&	3	&	0.002	&		&		&		&		&	2.4	\\
	&	V	&	3	&	0.003	&		&		&		&		&	2.4	\\
2009 Mar 22	&	B	&	3	&	0.004	&		&		&		&		&	2.4	\\
	&	R	&	5	&	0.005	&		&		&		&		&	2.4	\\
	&	V	&	5	&	0.003	&		&		&		&		&	2.4	\\
2009 Mar 23	&	B	&	5	&	0.007	&		&		&		&		&	2.4	\\
	&	R	&	4	&	0.002	&		&		&		&		&	2.4	\\
	&	V	&	4	&	0.001	&		&		&		&		&	2.4	\\
2009 Mar 26	&	B	&	5	&	0.013	&		&		&		&		&	2.4	\\
	&	R	&	5	&	0.006	&		&		&		&		&	2.4	\\
	&	V	&	5	&	0.003	&		&		&		&		&	2.4	\\
2009 Apr 14	&	I	&	61	&	0.011	&	0.6	&	1.84	&	2.3	&	2.39	&	1.02	\\
2009 Apr 16	&	I	&	20	&	0.007	&	2.41	&	3.03	&	0.78	&	4.7	&	1.02	\\
	&	R	&	20	&	0.023	&	0.15	&	3.03	&	0.83	&	4.7	&	1.02	\\
2009 May 11	&	I	&	30	&	0.009	&	1.32	&	2.42	&	2.43	&	3.46	&	1.02	\\
2009 May 16	&	I	&	7	&	0.02	&		&		&		&		&	1.02	\\
2009 May 17	&	I	&	75	&	0.013	&	0.6	&	1.75	&	9.39	&	2.18	&	1.02	\\
2010 Feb 21	&	I	&	5	&	0.004	&		&		&		&		&	1.02	\\
	&	R	&	5	&	0.009	&		&		&		&		&	1.02	\\
	&	V	&	5	&	0.012	&		&		&		&		&	1.02	\\
2010 Feb 22	&	I	&	4	&	0.003	&		&		&		&		&	1.02	\\
	&	R	&	1	&		&		&		&		&		&	1.02	\\
	&	V	&	5	&	0.011	&		&		&		&		&	1.02	\\
2010 Feb 26	&	B	&	10	&	0.008	&	0.95	&	5.35	&	1.31	&	9.55	&	2.4	\\
	&	I	&	8	&	0.002	&		&		&		&		&	2.4	\\
	&	R	&	7	&	0.001	&		&		&		&		&	2.4	\\
	&	V	&	10	&	0.006	&	0.18	&	5.35	&	18.37	&	9.55	&	2.4	\\
2010 Feb 28	&	B	&	15	&	0.143	&	0.01	&	3.7	&	0.42	&	5.99	&	2.4	\\
	&	I	&	15	&	0.018	&	1.84	&	3.7	&	1.16	&	5.99	&	2.4	\\
	&	R	&	15	&	0.015	&	0.17	&	3.7	&	1.75	&	5.99	&	2.4	\\
	&	V	&	15	&	0.025	&	0.22	&	3.7	&	1.9	&	5.99	&	2.4	\\
2010 Mar 11	&	B	&	1	&		&		&		&		&		&	2.4	\\
	&	I	&	1	&		&		&		&		&		&	2.4	\\
	&	R	&	1	&		&		&		&		&		&	2.4	\\
	&	V	&	1	&		&		&		&		&		&	2.4	\\
2010 May 04	&	R	&	19	&	0.007	&	0.92	&	3.13	&	0.95	&	4.86	&	1.02	\\
	&	V	&	18	&	0.01	&	0.37	&	3.24	&	0.77	&	5.06	&	1.02	\\
2010 May 15	&	R	&	88	&	0.005	&	5.14	&	1.69	&	4.37	&	2.08	&	1.02	\\
2010 May 16	&	R	&	88	&	0.007	&	1.03	&	1.69	&	1.67	&	2.08	&	1.02	\\
2010 May 17	&	R	&	80	&	0.005	&	5.31	&	1.75	&	1.75	&	2.17	&	1.02	\\
2010 May 18	&	R	&	70	&	0.005	&	1.12	&	1.79	&	2.26	&	2.27	&	1.02	\\
2011 May 07	&	R	&	60	&	0.007	&	1.44	&	1.93	&	1.18	&	2.39	&	1.02	\\
2011 May 09	&	R	&	63	&	0.007	&	2.11	&	1.84	&	2.15	&	2.37	&	1.02	\\
2011 May 10	&	R	&	38	&	0.012	&	0.47	&	2.21	&	0.69	&	3.02	&	1.02	\\
2011 Aug 21	&	B	&	5	&	0.046	&		&		&		&		&	1.02	\\
	&	I	&	5	&	0.013	&		&		&		&		&	1.02	\\
	&	R	&	5	&	0.019	&		&		&		&		&	1.02	\\
	&	V	&	5	&	0.019	&		&		&		&		&	1.02	\\
2011 Aug 22	&	B	&	5	&	0.054	&		&		&		&		&	1.02	\\
	&	I	&	5	&	0.065	&		&		&		&		&	1.02	\\
	&	R	&	4	&	0.06	&		&		&		&		&	1.02	\\
	&	V	&	5	&	0.075	&		&		&		&		&	1.02	\\
2012 Feb 27	&	I	&	5	&	0.009	&		&		&		&		&	1.02	\\
	&	R	&	5	&	0.012	&		&		&		&		&	1.02	\\
	&	V	&	5	&	0.021	&		&		&		&		&	1.02	\\
2012 Feb 28	&	I	&	5	&	0.011	&		&		&		&		&	1.02	\\
	&	R	&	5	&	0.023	&		&		&		&		&	1.02	\\
	&	V	&	4	&	0.011	&		&		&		&		&	1.02	\\
2012 Apr 02	&	I	&	4	&	0.004	&		&		&		&		&	1.02	\\
	&	R	&	5	&	0.026	&		&		&		&		&	1.02	\\
	&	V	&	4	&	0.039	&		&		&		&		&	1.02	\\
2012 Apr 29	&	B	&	4	&	0.056	&		&		&		&		&	1.02	\\
	&	I	&	4	&	0.009	&		&		&		&		&	1.02	\\
	&	R	&	4	&	0.009	&		&		&		&		&	1.02	\\
	&	V	&	4	&	0.019	&		&		&		&		&	1.02	\\
2012 May 01	&	I	&	5	&	0.007	&		&		&		&		&	1.02	\\
	&	R	&	5	&	0.005	&		&		&		&		&	1.02	\\
	&	V	&	5	&	0.028	&		&		&		&		&	1.02	\\
2012 May 03	&	B	&	4	&	0.03	&		&		&		&		&	1.02	\\
	&	I	&	4	&	0.022	&		&		&		&		&	1.02	\\
	&	R	&	4	&	0.011	&		&		&		&		&	1.02	\\
	&	V	&	4	&	0.021	&		&		&		&		&	1.02	\\
2012 May 11	&	I	&	5	&	0.011	&		&		&		&		&	1.02	\\
	&	R	&	5	&	0.014	&		&		&		&		&	1.02	\\
	&	V	&	5	&	0.025	&		&		&		&		&	1.02	\\
2012 May 13	&	V	&	1	&		&		&		&		&		&	1.02	\\
2012 May 16	&	R	&	31	&	0.02	&	0.26	&	2.39	&	1.11	&	3.4	&	1.02	\\
2012 May 17	&	R	&	26	&	0.015	&	0.58	&	2.6	&	2.13	&	3.84	&	1.02	\\

\enddata
\tablecomments{Column 1: observation dates; Column 2: observation filters; Column 3: observation numbers;
Column 4: observational errors estimated by standard star 1 and star 6; Column 5: the $F$ values in the $F$
test for the observation data; Column 6: $F$(99) is the critical $F$ value at a 99\% confidence level;
Column 7: the ANOVA values in the ANOVA test for the observation data; Column 8: ANOVA(99) is the critical
ANOVA value at a 99\% confidence level; Column 9: 1.02 and 2.4 represent 1.02 m and 2.4 m telescopes, respectively.
}
\end{deluxetable}

\clearpage

\begin{deluxetable}{lccccccccccc}
  \tablecolumns{12}
  \setlength{\tabcolsep}{3pt}
  \tablewidth{0pc}
  \tablecaption{The observational data for Mrk 501}
  \tabletypesize{\scriptsize}
  \tablehead{
  \multicolumn{3}{c}{$B$}           &
  \multicolumn{3}{c}{$I$}           &
  \multicolumn{3}{c}{$R$}           &
  \multicolumn{3}{c}{$V$}            \\
  \cline{1-3} \cline{4-6} \cline{7-9} \cline{10-12}

  \colhead{MJD - 50000}                                        &
  \colhead{Mag}                                          &
   \colhead{FWHM}                                       &

  \colhead{MJD - 50000}                                        &
  \colhead{Mag}                                             &
  \colhead{FWHM}                                       &

  \colhead{MJD - 50000}                                      &
  \colhead{Mag}                                           &
    \colhead{FWHM}                                       &

  \colhead{MJD - 50000}                                 &
  \colhead{Mag}                                       &
  \colhead{FWHM}
}
\startdata
 3465.856933  &  14.459  &    3.39	&	 3465.847570  &  12.579  &    2.65	&	 3465.865243  &  12.942  &    3.36	&	 3465.853114  &  13.382  &    3.24				\\
 3465.872859  &  14.431  &    3.39	&	 3465.862686  &  12.576  &    2.66	&	 3465.881111  &  12.938  &    3.27	&	 3465.868264  &  13.361  &    3.15				\\
 3465.887859  &  14.450  &    4.45	&	 3465.878669  &  12.573  &    2.48	&	 3465.895648  &  12.943  &    3.26	&	 3465.883577  &  13.368  &    3.39				\\
 3465.902095  &  14.457  &    3.30	&	 3465.893345  &  12.586  &    2.77	&	 3465.909653  &  12.942  &    3.20	&	 3465.898079  &  13.368  &    3.57				\\
 3466.806713  &  14.602  &    2.49	&	 3465.907292  &  12.587  &    2.37	&	 3466.800787  &  13.146  &    1.85	&	 3465.912072  &  13.364  &    3.36				\\
 4205.897373  &  14.254  &    3.29	&	 3466.798473  &  12.459  &    1.69	&	 3466.813588  &  13.137  &    2.55	&	 3466.802917  &  13.579  &    2.31				\\
 4205.900324  &  14.254  &    3.18	&	 3466.811401  &  12.460  &    2.29	&	 3466.822234  &  13.132  &    2.28	&	 3466.815973  &  13.592  &    2.37				\\
 4229.711204  &  14.478  &    1.71	&	 3466.819977  &  12.467  &    2.22	&	 3466.831366  &  13.146  &    2.29	&	 3466.824537  &  13.583  &    2.58				\\
...&...&...&...&...&...&...&...&...&...&...&... \\
\enddata
\tablecomments{This table is available in its entirety in a machine-readable form in the online journal. A portion is shown here for
guidance regarding its form and content. FWHM is for the standard stars, and is in units of arc-seconds. Mag denotes magnitude.}
\end{deluxetable}

\clearpage

\begin{figure*}
  \begin{center}
   \includegraphics[width=0.45\textwidth]{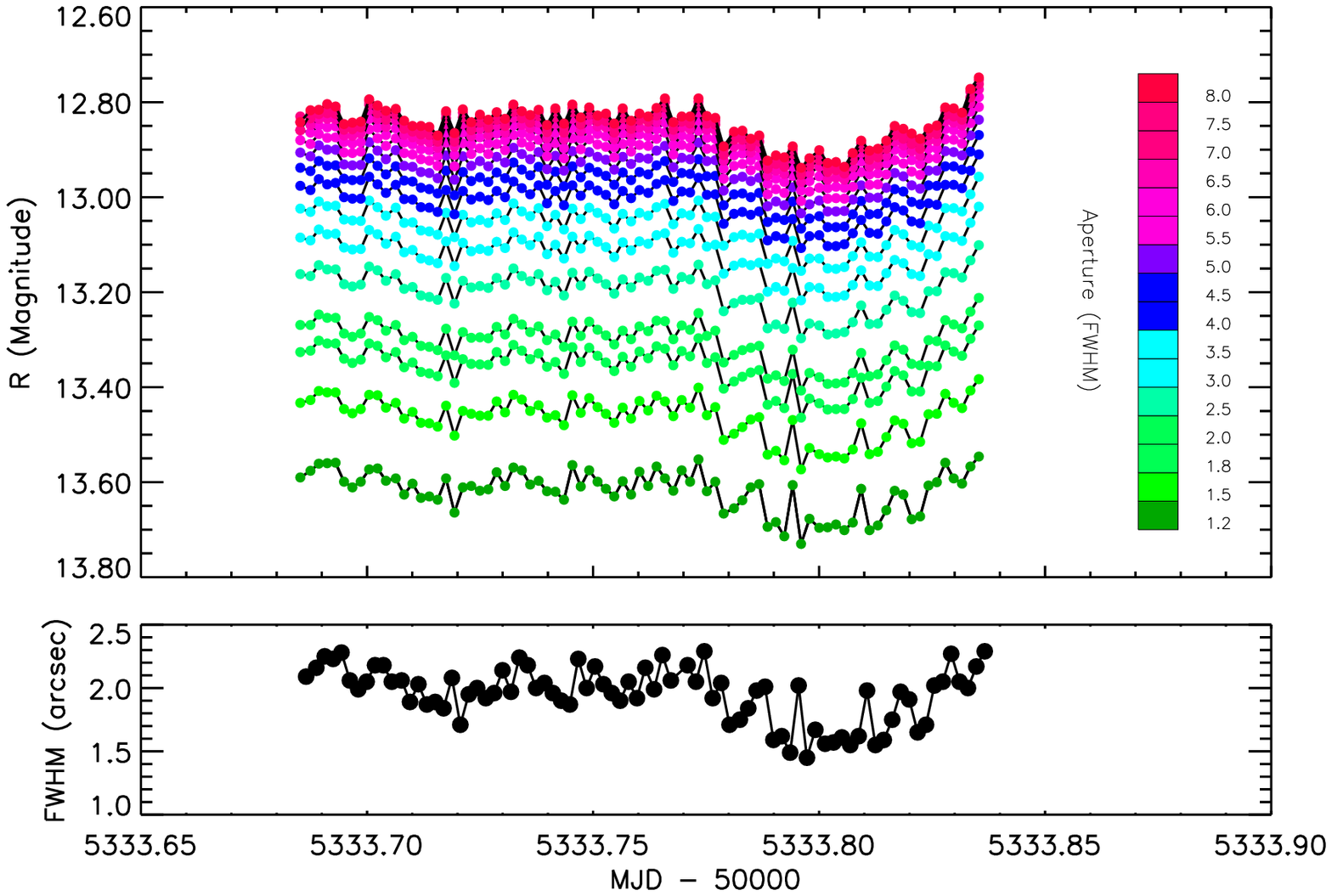}
   \includegraphics[width=0.45\textwidth]{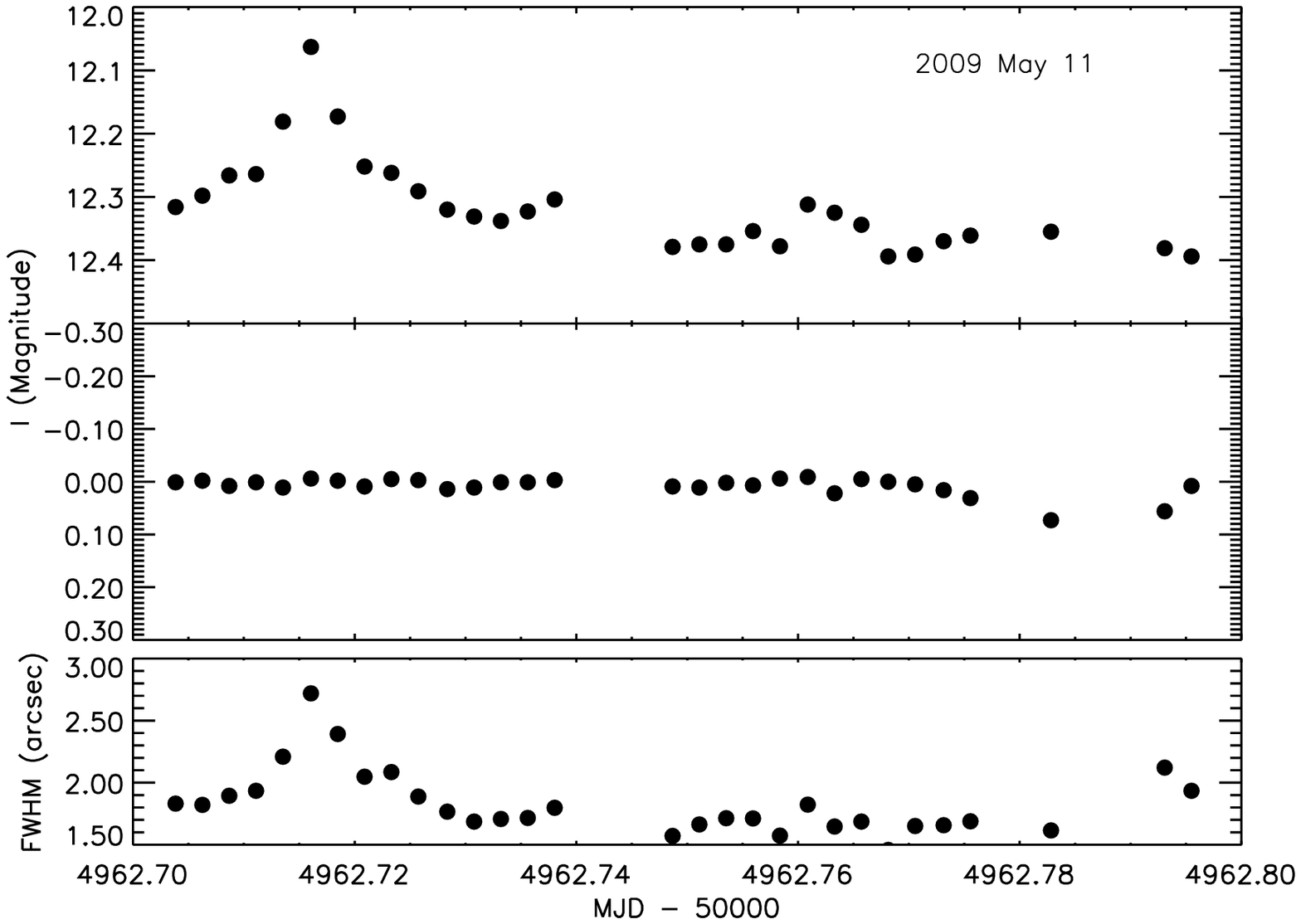}
  \end{center}
  \caption{Examples of the aperture radii test for the $R$ band on 2010 May 17 and the $I$ band on 2009 May 11. In the left panels, the
  color bar shows the different aperture in units of FWHM, and the lower panel shows the FWHM variability. In the right panels, the top
  panel is the light curve as an aperture of 6.0 FWHMs is used, the middle panel is the differential variations of two comparison stars,
  and the bottom panel is the FWHM variability. }
  \label{fig1}
\end{figure*}

\begin{figure*}
 \begin{center}
  \includegraphics[angle=0,scale=0.45]{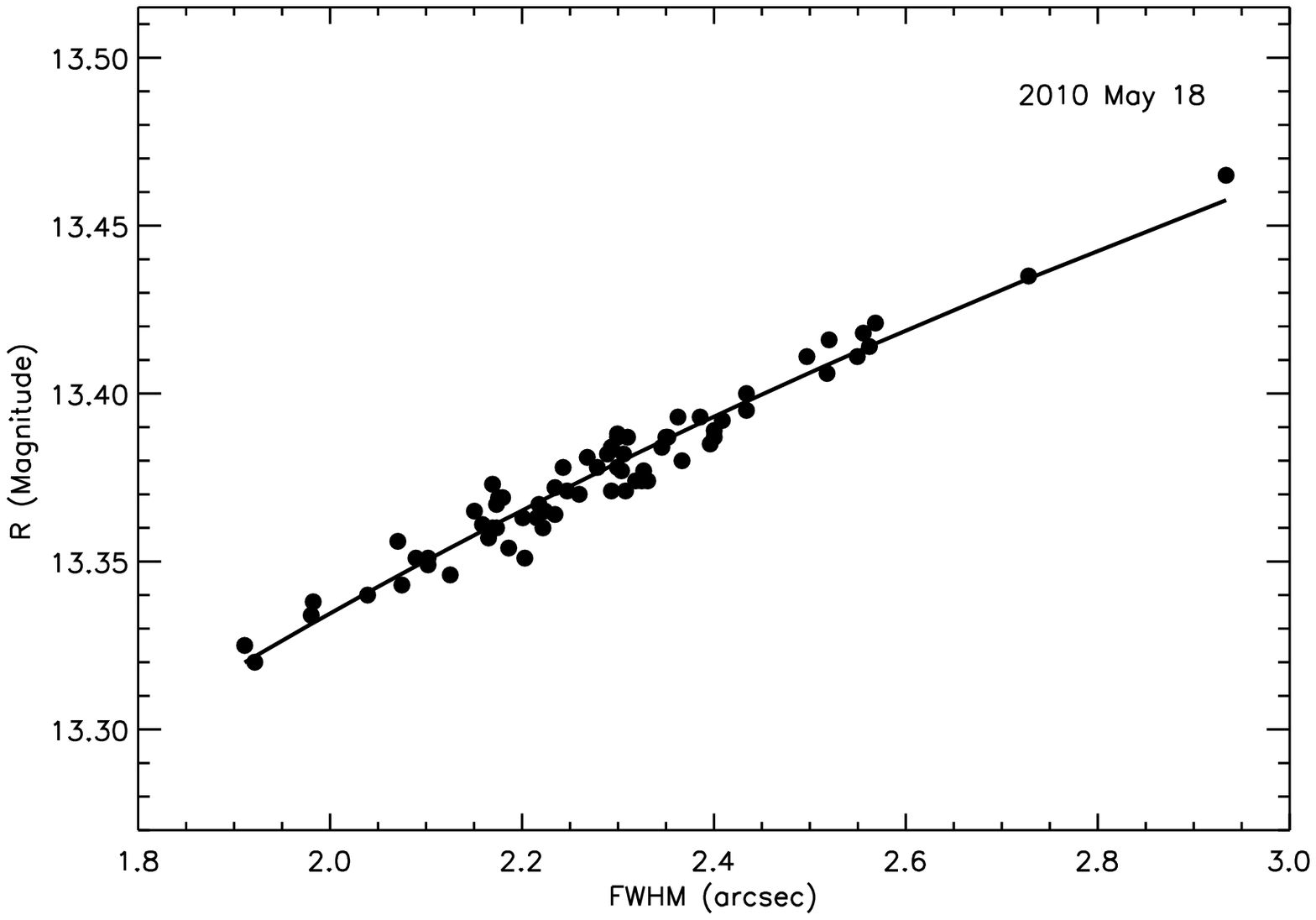}
  \includegraphics[angle=0,scale=0.45]{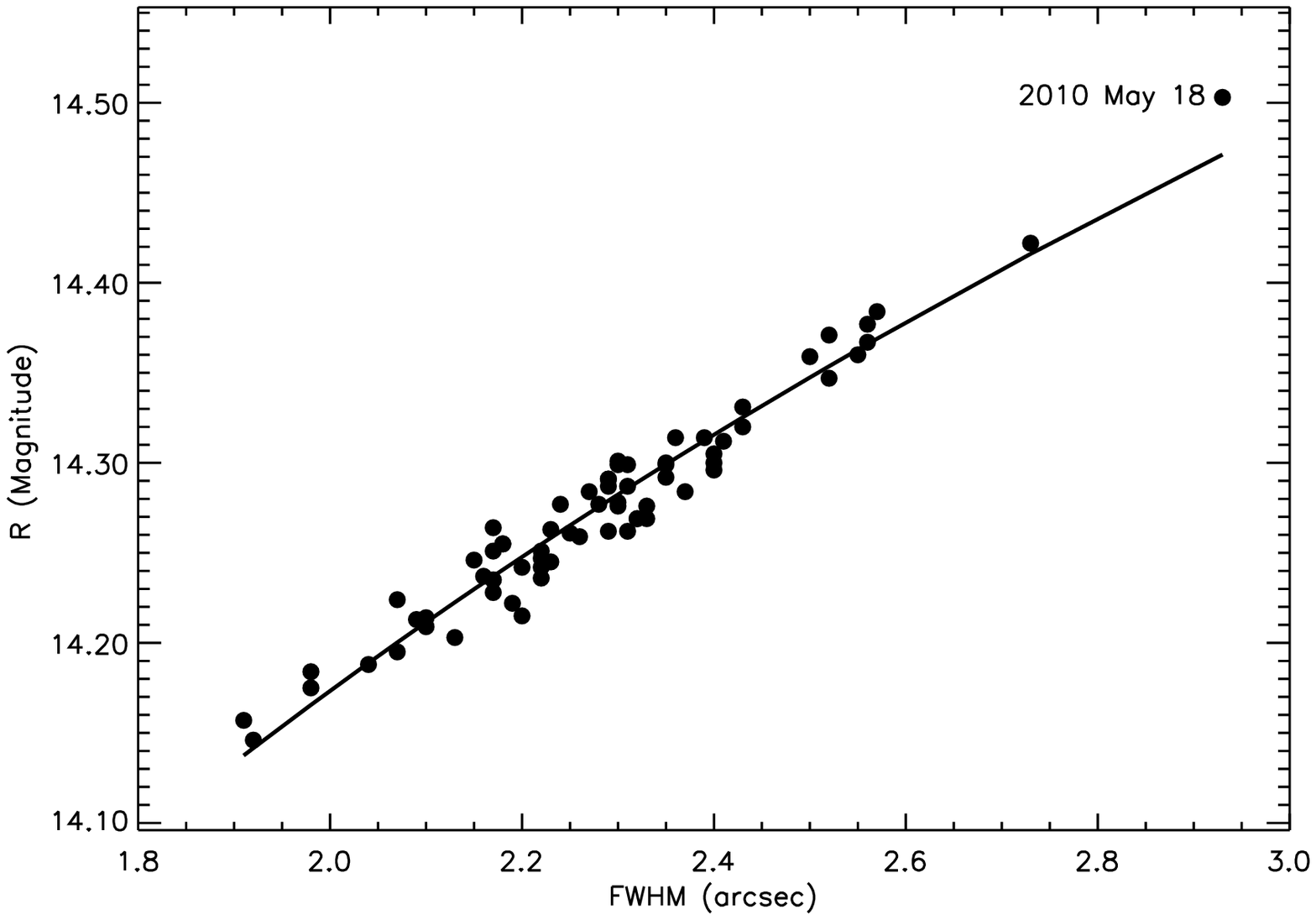}
 \end{center}
 \caption{Example of seeing influence on photometric results of Mrk 501 (i.e., the FWHM--magnitude relation). The left
  panel denotes the observational results on 2010 may 18. The right panel shows the host-subtracted results according to
  \citet{Ni07}. Black solid circles denote the observational results, and solid lines are the fit ones.}
  \label{fig2}
\end{figure*}

\begin{figure*}
  \begin{center}
   \includegraphics[width=0.45\textwidth]{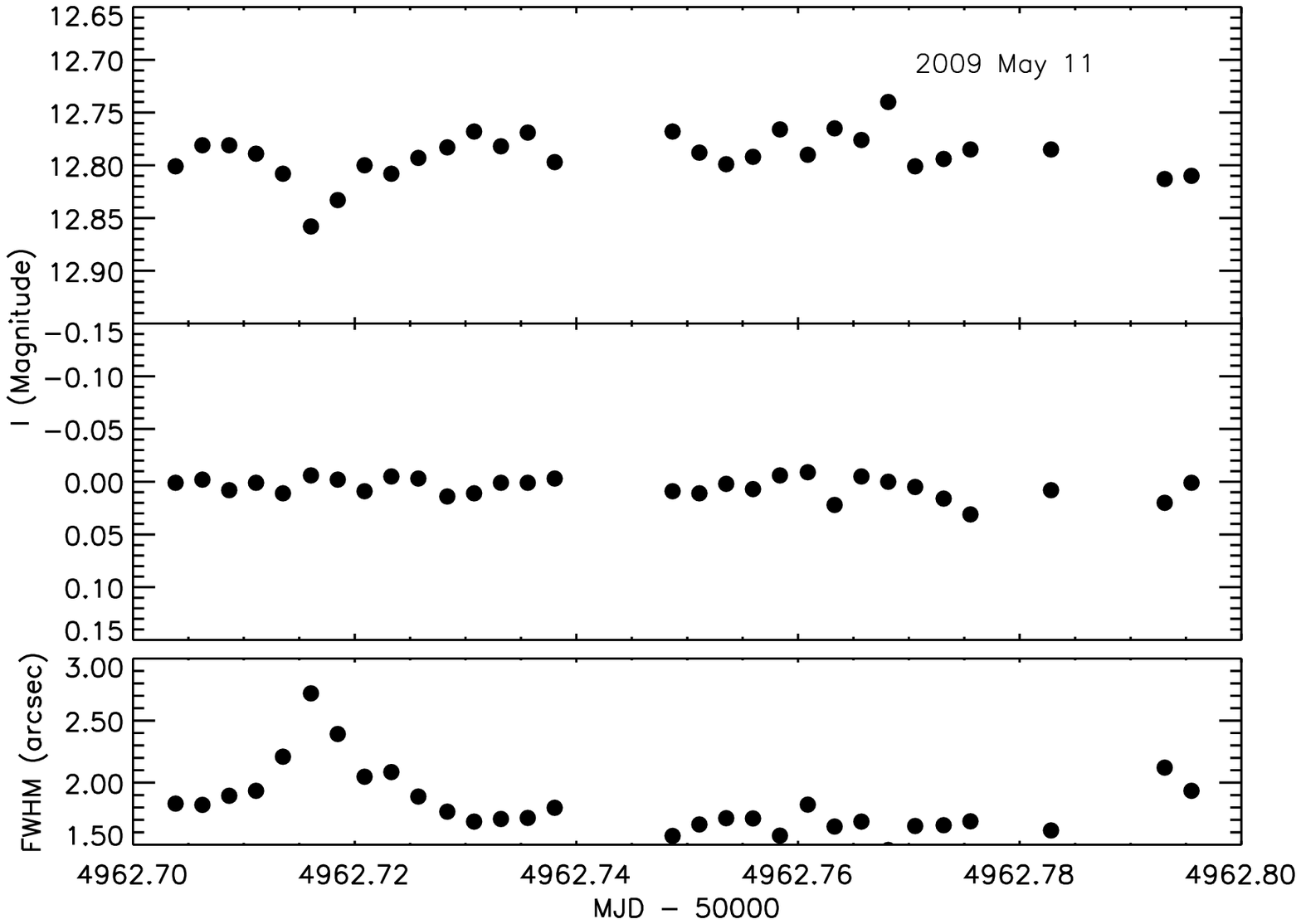}
   \includegraphics[width=0.45\textwidth]{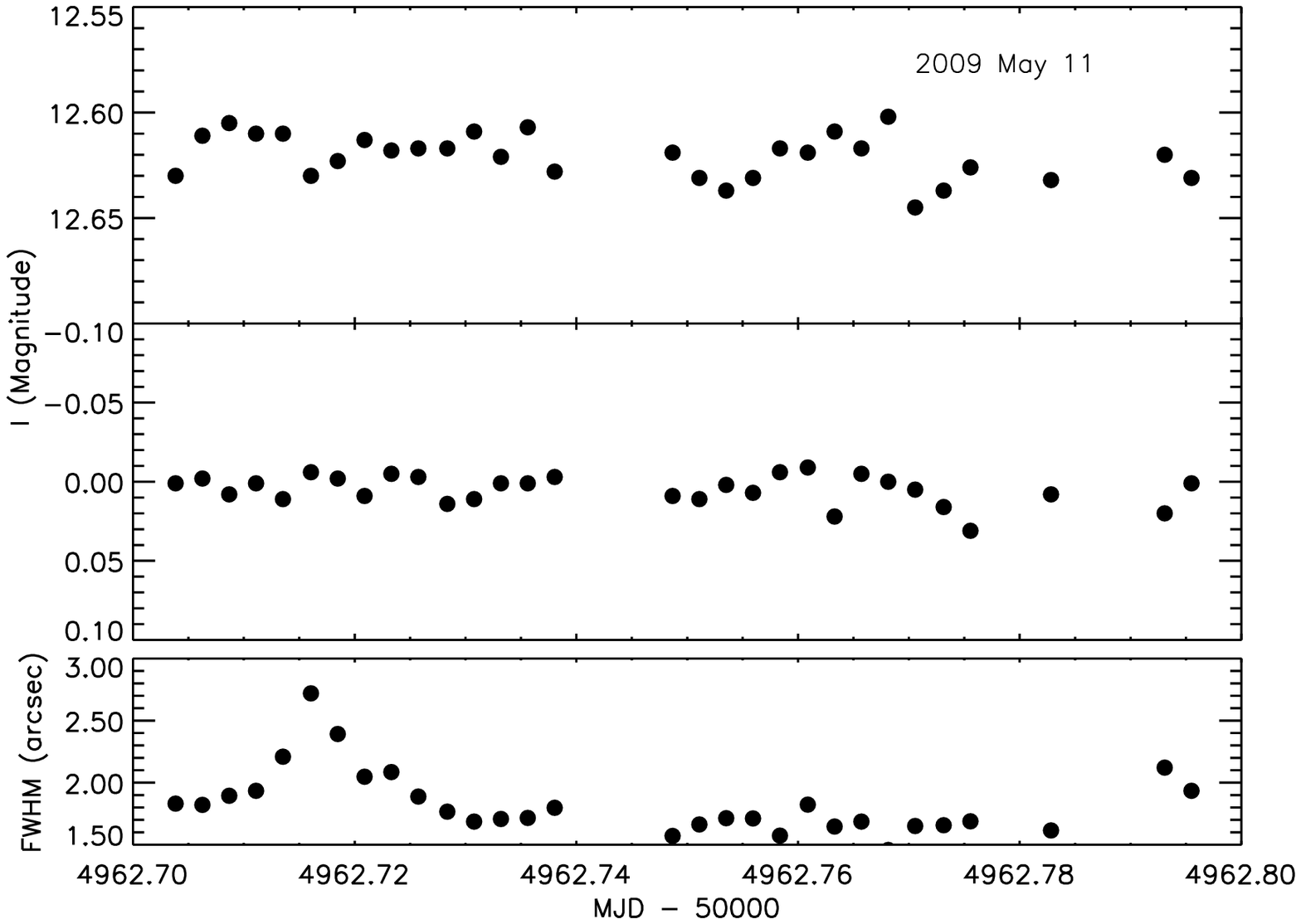}
  \end{center}
  \caption{Example of the correcting seeing effect for the $I$ band on 2009 May 11. The curves are the same as in the right panel
   of Figure 1, except for a fixed aperture. The left panel denotes the results without correcting the seeing effect, and the right panel
   denotes the corrected results. }
  \label{fig3}
\end{figure*}

\begin{figure*}
 \begin{center}
  \includegraphics[angle=0,scale=0.45]{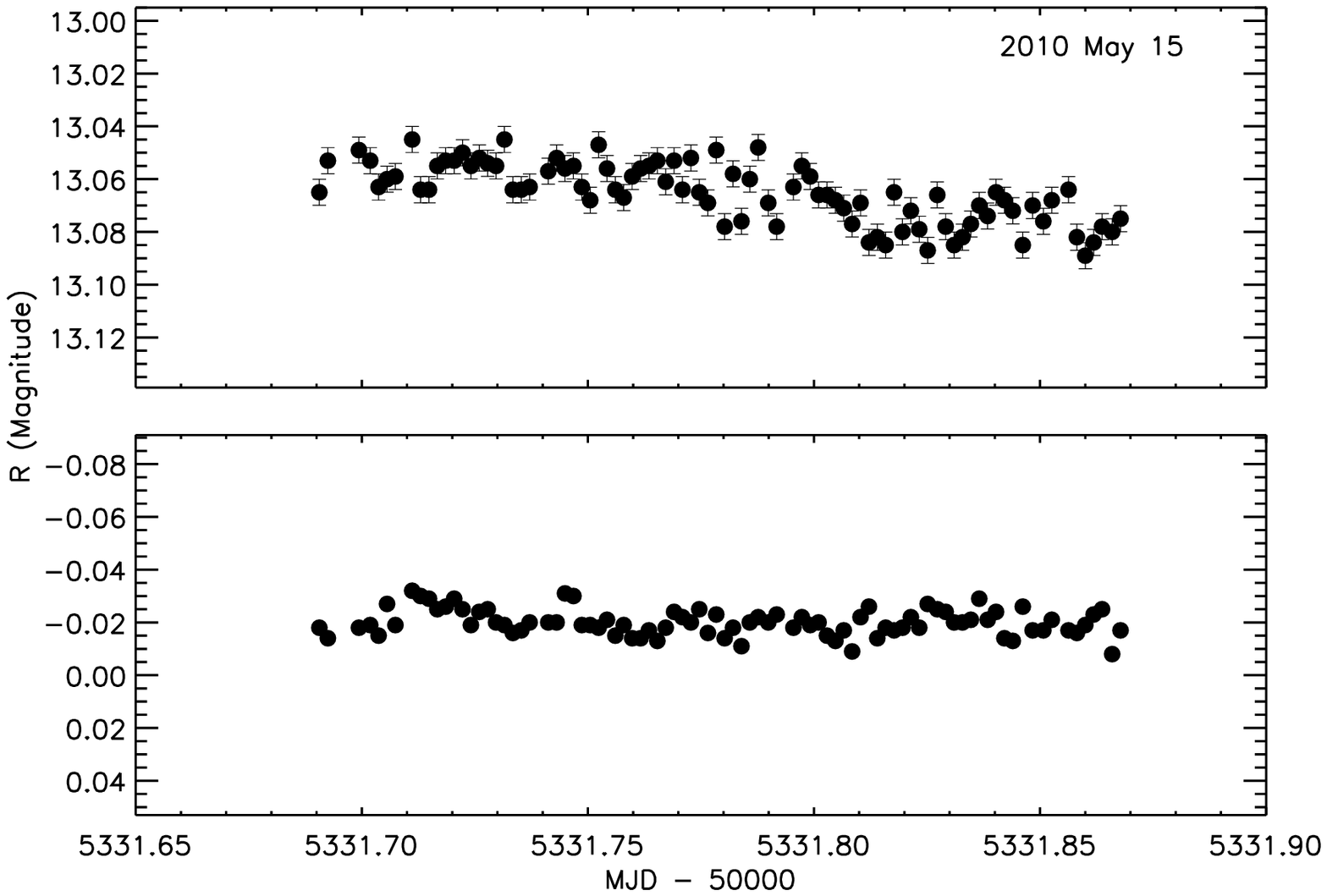}
  \includegraphics[angle=0,scale=0.45]{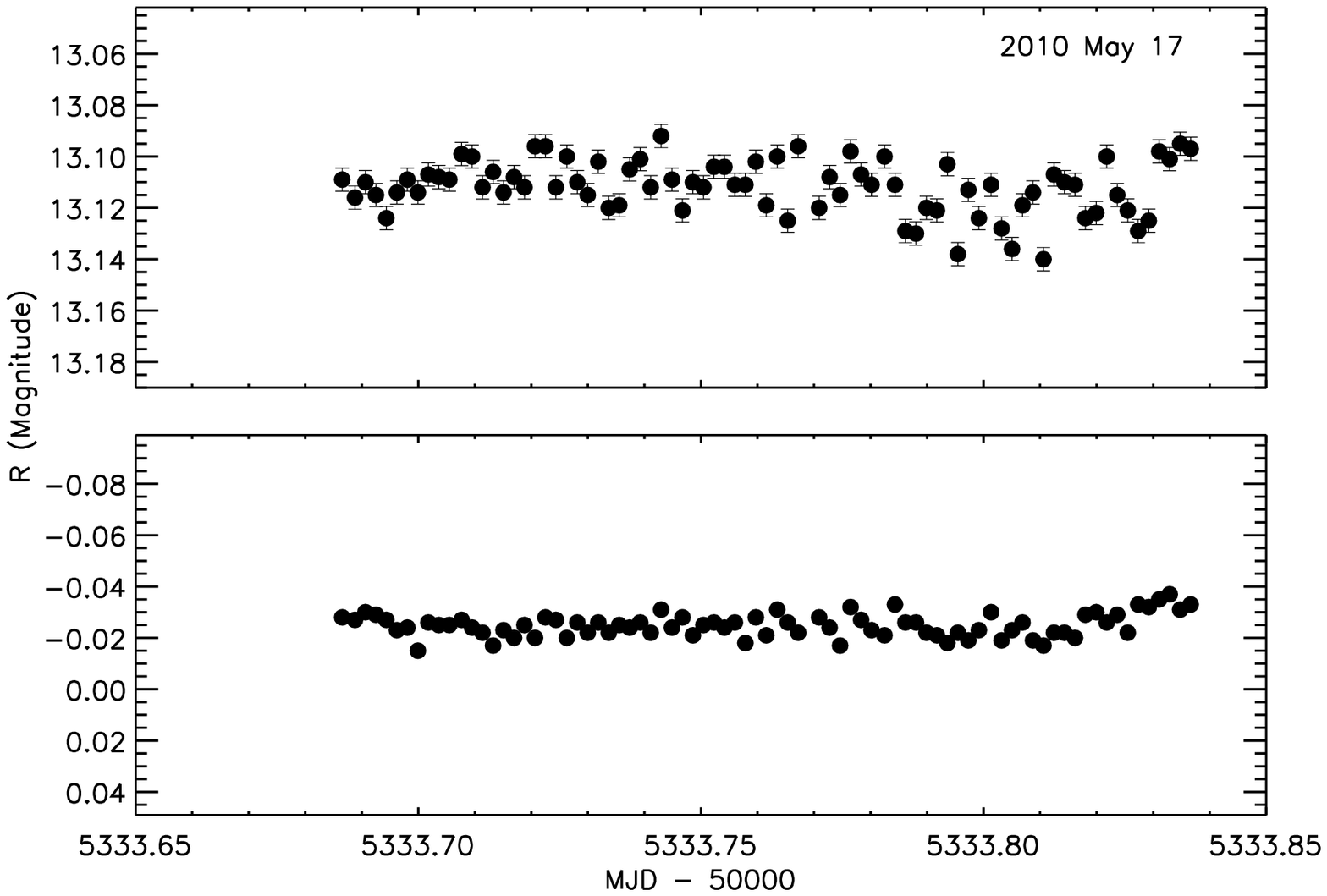}
  \end{center}
 \caption{Examples of the possible IDVs for Mrk 501. For each night, the top panel is the light curve of Mrk 501, and
  the bottom panel is the differential variations of comparison stars 1 and 6.}
  \label{fig4}
\end{figure*}

\begin{figure*}
 \begin{center}
  \includegraphics[angle=0,scale=0.4]{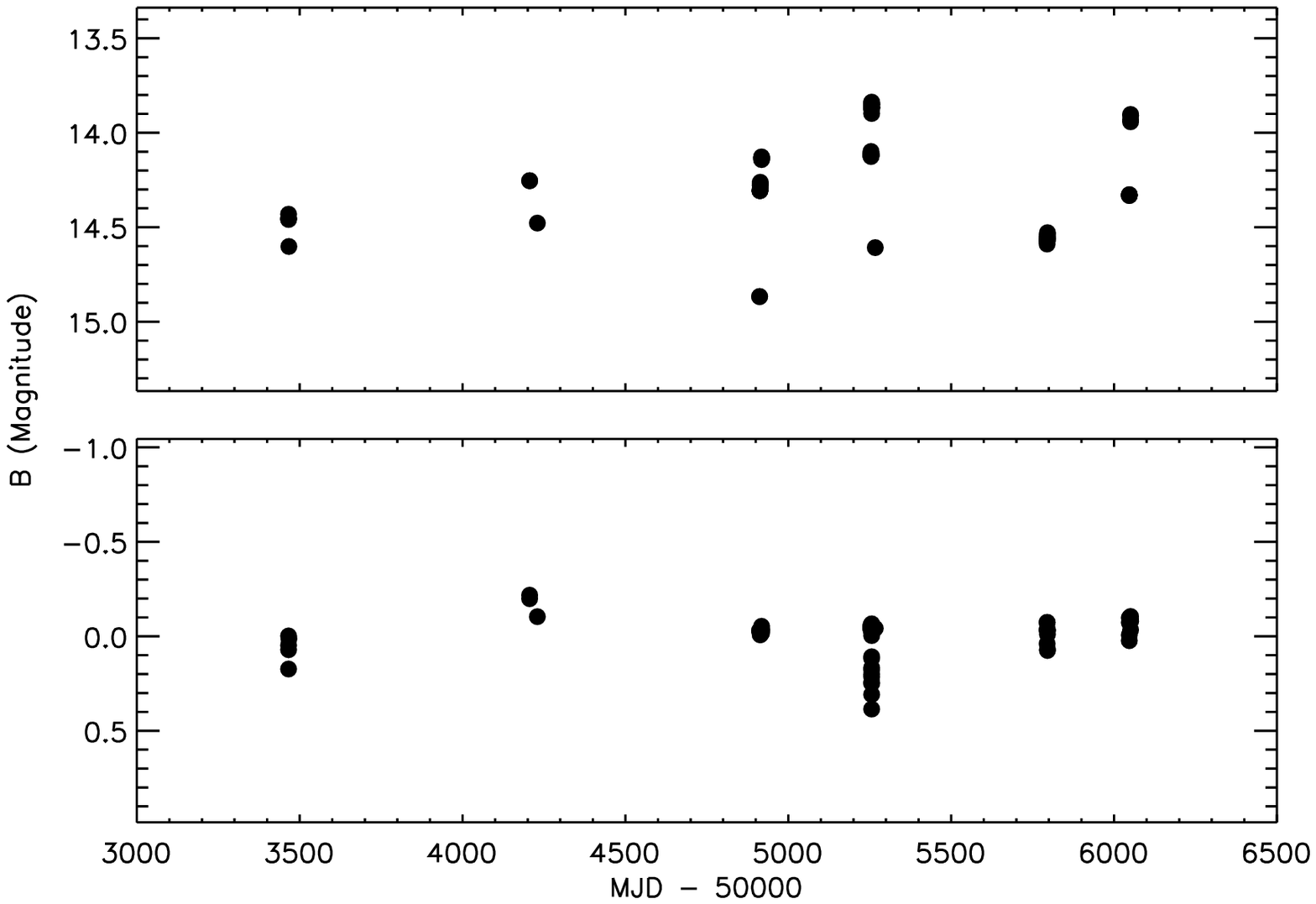}
  \includegraphics[angle=0,scale=0.4]{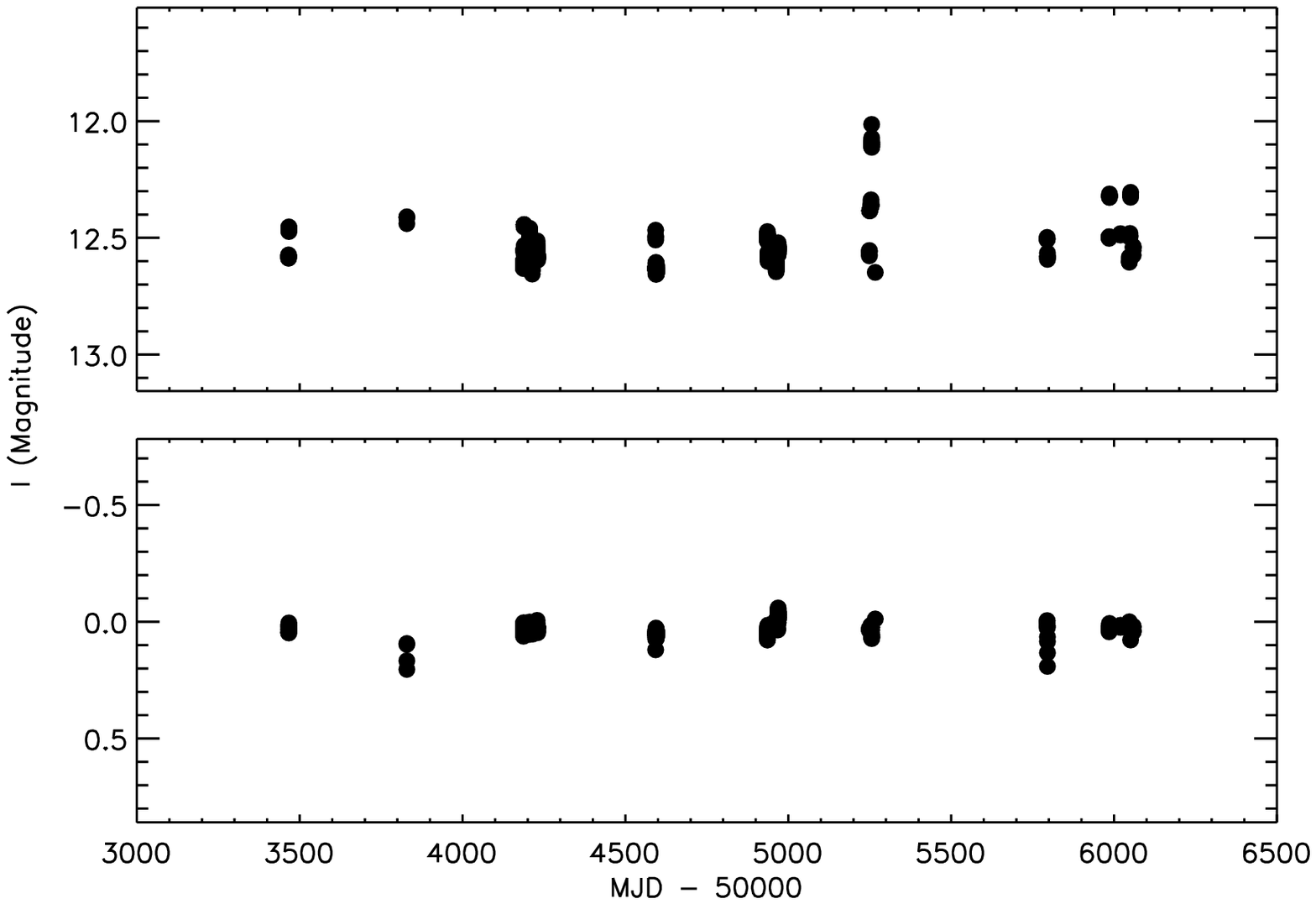}
  \includegraphics[angle=0,scale=0.4]{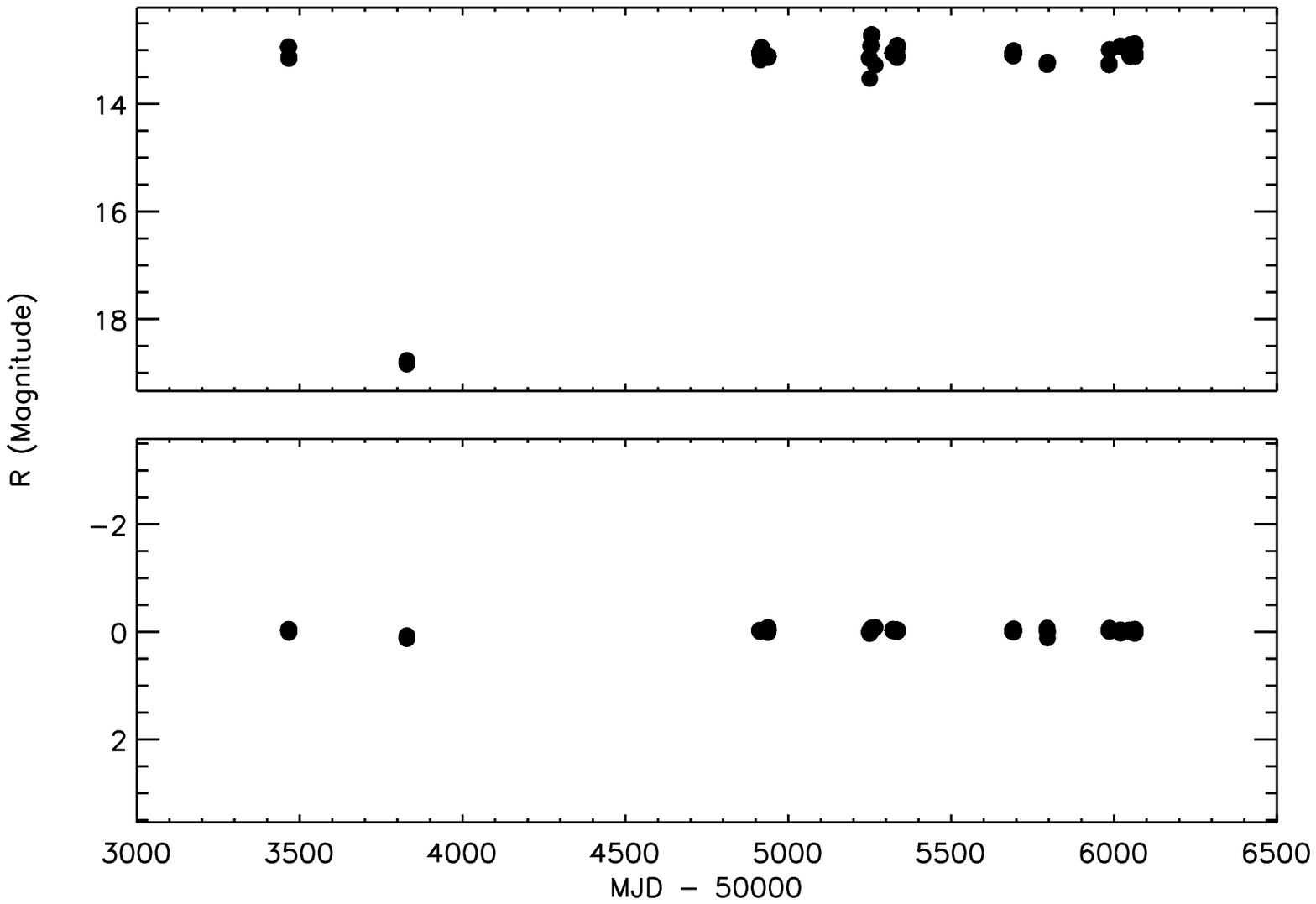}
  \includegraphics[angle=0,scale=0.4]{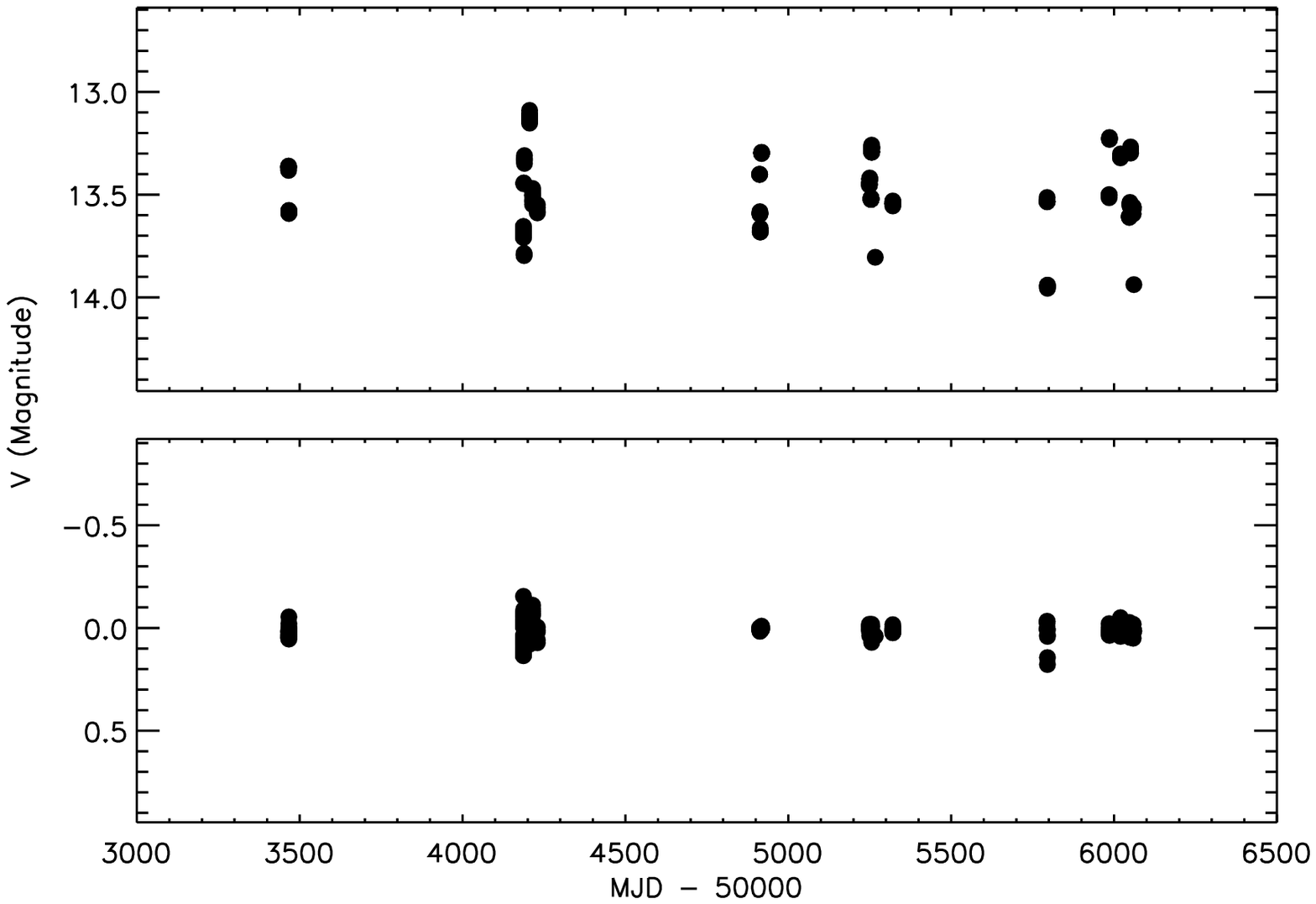}
 \end{center}
 \caption{Long term light curves in the $B$, $I$, $R$, and $V$ bands for Mrk 501. For each band, the top panel is
  the light curve and the bottom panel is the differential variations of two comparison stars 1 and 6.}
  \label{fig5}
\end{figure*}

\begin{figure*}
 \begin{center}
  \includegraphics[angle=-90,scale=0.45]{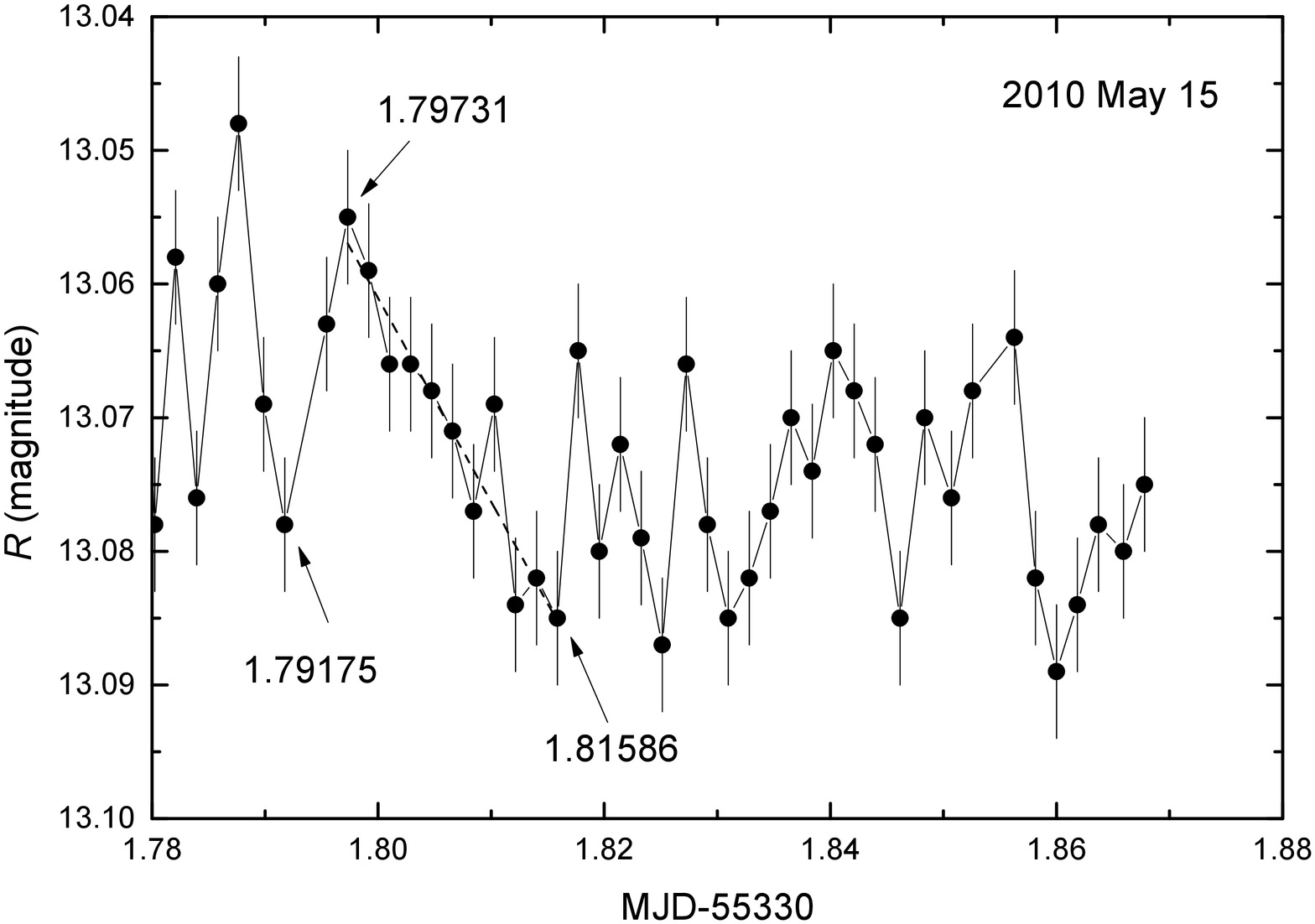}
 \end{center}
 \caption{A possible IDV in one night for Mrk 501. The numbers in panel are the corresponding times of data points denoted
  by the arrows. The dashed line is the best linear fitting to the 11 data points with the $y$ errors [$y=10.32(\pm 0.32)+1.52(\pm 0.18)x$].}
  \label{fig6}
\end{figure*}

\begin{figure*}
 \begin{center}
  \includegraphics[angle=-90,scale=0.45]{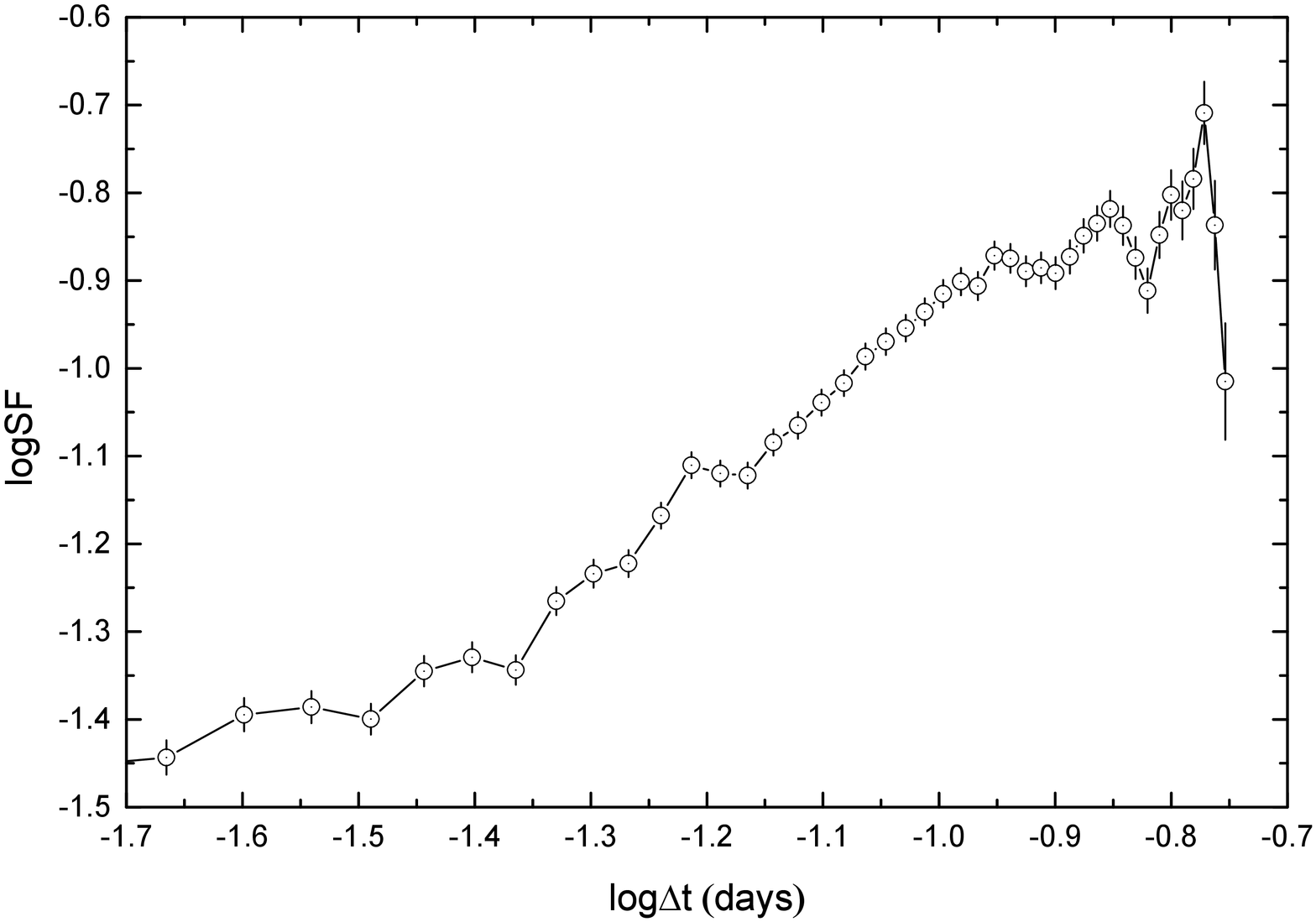}
 \end{center}
 \caption{\textbf{Structure Function for the light curve on 2010 May 15.} }
  \label{fig7}
\end{figure*}

\end{document}